\documentclass{article}

\usepackage[english]{babel}
\usepackage{amsmath,amssymb} 
\usepackage{txfonts}
\usepackage{subfigure}
\usepackage{graphicx}
\usepackage{ifpdf}
\ifpdf
  \usepackage[dvipdfm,bookmarks,colorlinks,hyperindex]{hyperref}
\else
  \usepackage[hypertex,colorlinks,hyperindex]{hyperref}
\fi

\newcommand{\e}{\mathrm{e}}
\newcommand{\erf}{\mathrm{erf}}

\begin{document}

\title{A direct Numerov sixth order numerical scheme to accurately
    solve the unidimensional Poisson equation with Dirichlet boundary
    conditions}
\author{Esmerindo Bernardes\\
  Departamento de F{\'{\i}}sica e Ci{\^{e}}ncia dos Materiais\\
  Instituto de F{\'{\i}}sica de S{\~{a}}o Carlos\\
  Universidade de S{\~{a}}o Paulo\\
  Av. do Trabalhador S{\~{a}}o-carlense, 400 CP 369\\
  13560.970 S{\~{a}}o Carlos, SP, Brasil}

\date{\today}
\maketitle

\begin{abstract}
  In this article, we present an analytical direct method, based on a
  Numerov three-point scheme, which is sixth order accurate and has
  a linear execution time on the grid dimension, to solve the discrete
  one-dimensional Poisson equation with Dirichlet boundary
  conditions. Our results should improve numerical codes used mainly
  in self-consistent calculations in solid state physics.
\end{abstract}

\section{Introduction}
The one-dimensional Poisson equation,
\begin{equation}
  \label{eq:Poisson1D}
  \frac{d^{2}\phi}{dx^{2}} = -\rho,\quad a\leq x\leq b,
\end{equation}
with Dirichlet boundary conditions,
\begin{equation}
  \label{eq:Dirichlet1D}
  \phi(a)=c_{1},\quad \phi(b)=c_{2},
\end{equation}
plays an important role in many branches of science. Particularly, the
Poisson equation \eqref{eq:Poisson1D} is essential in self-consistent
calculations in solid state physics~\cite{johnson2001}. In general, we
have to solve it numerically many times. Therefore, is vital to have
the fastest and the most accurate numerical scheme to solve it. In
this article, we present a very efficient direct method, based on a
Numerov~\cite{rasmn84bvn1924,jcp1jmb1967,cma0rpa2001}
sixth order numerical scheme, to solve the Poisson equation
\eqref{eq:Poisson1D} numerically. Because of its efficiency and
simplicity, this new method can be used as a canonical numerical
scheme to accurately solve the one-dimensional Poisson equation.

This article is organized as follows. Our numerical scheme is
presented in Section~\ref{sec:numerov}. Its linearization, together
with a few discussions, are presented in Section~\ref{sec:discus}. Our
conclusions are presented in Section~\ref{sec:conclus}.

\section{The Numerov scheme}
\label{sec:numerov}
Let $\phi_{i}=\phi(x_{i})$ represents the solution of
\eqref{eq:Poisson1D} at the $i$-th point, $x_{i}=a+(i-1)h$, of an
equally spaced net of step $h=(b-a)/N$ and dimension $N+1$. Let also
$\phi_{i}^{(k)}$ represents the $k$-th derivative evaluated at the
same point $x_{i}$. Then we can evaluate the solution $\phi$ at the
nearest neighborhood points $x_{i\pm 1}$ of $x_{i}$ using Taylor
series~\cite{thomas1995},
\begin{equation}
  \label{eq:taylor}
  \phi_{i\pm 1} = \phi(x_{i}\pm h) = \phi_{i} \pm \phi_{i}^{(1)}\, h +
  \frac{1}{2} \phi_{i}^{(2)}\, h^{2} \pm \mathcal{O}(h^{3}).
\end{equation}

The basic idea in the Numerov approach is to eliminate the fourth
order derivative in the expression
\begin{equation}
  \alpha_{1}A_{1}+\alpha_{2}A_{2} = \alpha_{1}\phi_{i} -
  \bigl(\frac{h^{2}}{2}\alpha_{1}+\alpha_{2}\bigr)\rho_{i} + 
  \bigl(\frac{h^{4}}{24}\alpha_{1} +
  \frac{h^{2}}{2}\alpha_{2}\bigr)\phi_{i}^{(4)} + \mathcal{O}(h^{6}),
\end{equation}
where
\begin{align}
  \label{eq:A1}
  A_{1} &= \frac{1}{2}(\phi_{i+1}+\phi_{i-1}) = \phi_{i} -
  \frac{h^{2}}{2}\rho_{i} + \frac{h^{4}}{24}\phi_{i}^{(4)} + \cdots, \\
  A_{2} &= \frac{1}{2}(\phi_{i+1}^{(2)}+\phi_{i-1}^{(2)}) = -
  \frac{1}{2}(\rho_{i+1}+\rho_{i-1})
  = -\rho_{i} + \frac{h^{2}}{2}\phi_{i}^{(4)} + \cdots,
  \label{eq:A2}
\end{align}
to obtain the sixth order three-point numerical scheme
\begin{equation}
  \label{eq:Num}
  \phi_{i\pm 1} = 2\phi_{i} - \phi_{i\mp 1} - \frac{h^{2}}{12}\bigl(
  \rho_{i+1} + 10\rho_{i} + \rho_{i-1} \bigr), 
\end{equation}
where we chose $\alpha_{1}=1$ and, consequently,
$\alpha_{2}=-h^{2}/12$. In a similar way, we can eliminate the third
order derivative from
\begin{equation}
  \beta_{1}B_{1}+\beta_{2}B_{2} = h\beta_{1}\phi_{i}^{(1)} +
  \bigl(\frac{h^{3}}{6}\beta_{1}+h\beta_{2}\bigr)\rho_{i}^{(3)} + 
  \mathcal{O}(h^{5}),
\end{equation}
where
\begin{align}
  \label{eq:B1}
  B_{1} &= \frac{1}{2}(\phi_{i+1}-\phi_{i-1}) = h\phi_{i}^{(1)} +
  \frac{h^{3}}{6}\phi_{i}^{(3)} + \cdots, \\
  B_{2} &= \frac{1}{2}(\phi_{i+1}^{(2)}-\phi_{i-1}^{(2)}) = -
  \frac{1}{2}(\rho_{i+1}-\rho_{i-1})
  = -h\rho_{i}^{(3)} + \cdots,
  \label{eq:B2}
\end{align}
to obtain the fifth order three-point numerical scheme
\begin{equation}
  \label{eq:Num}
  \phi_{i}^{(1)} = \frac{1}{2h}\bigl(\phi_{i+1} - \phi_{i-1}\bigr) +
  \frac{h^{2}}{6} \bigl( \rho_{i+1} - \rho_{i-1} \bigr), 
\end{equation}
for the first derivative of $\phi$, where we chose $\beta_{1}=1$ and,
consequently, $\beta_{2}=-h^{2}/6$.

So far, the three-point numerical scheme \eqref{eq:Num} is an
iterative method, i.e., given two informations, $\phi_{i-1}$ and
$\phi_{i}$, we can calculate $\phi_{i+1}$. One difficulty of this
iterative method is related with the Dirichlet boundary conditions
\eqref{eq:Dirichlet1D}: they are known only at end-points $x_{1}$ and
$x_{N+1}$. Thus, we can not initiate our iterative scheme
\eqref{eq:Num}. Fortunately, the recurrence relation in
\eqref{eq:Num} is linear with constant coefficients. These
two features imply we can find an unique solution to it,
\begin{equation}
  \label{eq:recur}
  \phi_{i} = (i-1)\phi_{2} - (i-2)\phi_{1} - \frac{h^{2}}{12}
  \sum_{j=3}^{i} (i+1-j)(\rho_{j} + 10\rho_{j-1} + \rho_{j-2}),
\end{equation}
where $\phi_{1}=c_{1}$ and $\phi_{2}$ must be expressed in terms of
$\phi_{N+1}=c_{2}$ (the Dirichlet boundary conditions),
\begin{equation}
  \label{eq:Phi2}
  \phi_{2} = \frac{1}{N}\phi_{N+1} + (1-\frac{1}{N})\phi_{1} +
  \frac{h^{2}}{12N} \sum_{j=3}^{N+1} (N+2-j)(\rho_{j} + 10\rho_{j-1} +
  \rho_{j-2}). 
\end{equation}
Now we have an analytical sixth order numerical scheme to solve
accurately the Poisson equation \eqref{eq:Poisson1D} with the
Dirichlet boundary conditions \eqref{eq:Dirichlet1D}. 

It should be mentioned that the analytical third order numerical
scheme presented by Hu and O'Connell~\cite{jpa29gyh1996}, making use
of tridiagonal matrices, can also be derived by the present approach
restricted to the third order,
\begin{equation}
  \label{eq:recurh}
  \phi_{i} = (i-1)\phi_{2} - (i-2)\phi_{1} - h^{2}
  \sum_{j=3}^{i} (i+1-j)\rho_{j-1},
\end{equation}
where
\begin{equation}
  \label{eq:Phi2h}
  \phi_{2} = \frac{1}{N}\phi_{N+1} + (1-\frac{1}{N})\phi_{1} +
  h^{2} \sum_{j=3}^{N+1} (N+2-j)\rho_{j-1}. 
\end{equation}

\section{Discussions}
\label{sec:discus}
Although we have found a very accurate analytical direct method to
solve the one-dimensional Poisson equation with Dirichlet boundary
conditions, namely, the sixth order Numerov scheme \eqref{eq:Num}, it
has one undesirable feature: its execution time is proportional to
the square of the grid dimension. Fortunately it can be linearized.
First, we create a vector $U$, whose components are the partial sums
$U_{i}=\rho_{i}+10\rho_{i-1}+\rho_{i-2}$ ($U_{1}=U_{2}=0$). Next, we
create a second vector $V$ with $V_{i}=V_{i-1}+U_{i}$ and
$V_{1}=V_{2}=0$. We also need a third vector $Y$ with $Y_{i}=iU_{i}$
and a fourth vector $Z$ with the complete sums $Z_{i}=Z_{i-1}+Y_{i}$.
Using these new vectors, our sixth order Numerov scheme \eqref{eq:Num}
can be rewritten as follows,
\begin{equation}
  \label{eq:recur2}
  \phi_{i} = (i-1)\phi_{2} - (i-2)\phi_{1} - \frac{h^{2}}{12}
  \bigl[(i+1)V_{i} - Z_{i}\bigr].
\end{equation}
This numerical scheme has now a linear execution time proportional to
five times the grid dimension $N+1$.

Let us use a Gaussian density, 
\begin{equation}
  \label{eq:densidade}
  \rho(x) = \e^{-x^{2}/4},
\end{equation}
to verify the accuracy and the efficiency of the non-linear numerical
scheme \eqref{eq:recur}, as well as the linear numerical scheme
\eqref{eq:recur2}. The solution for the Poisson equation
\eqref{eq:Poisson1D}, along with the boundary conditions
$\phi(-10)=\phi_{1}=1$ and $\phi(+10)=\phi_{N+1}=2$, is
\begin{equation}
  \label{eq:sol_exata}
  \phi(x) = \frac{x}{20}  - \sqrt{\pi}\,x\,\erf(x/2) - 2\e^{-x^{2}/4} +
  \frac{3}{2} + 10\sqrt{\pi}\,\erf(5) + 2\e^{-25},
\end{equation}
where $\erf(x)$ is the error function,
\begin{equation}
  \erf(x) = \frac{2}{\sqrt{\pi}} \int_{0}^{x}\e^{-t^2}\, dt.
\end{equation}

\begin{figure}[h]
  \centering
  \subfigure[Execution times]{\label{f1}\includegraphics[scale=0.5]{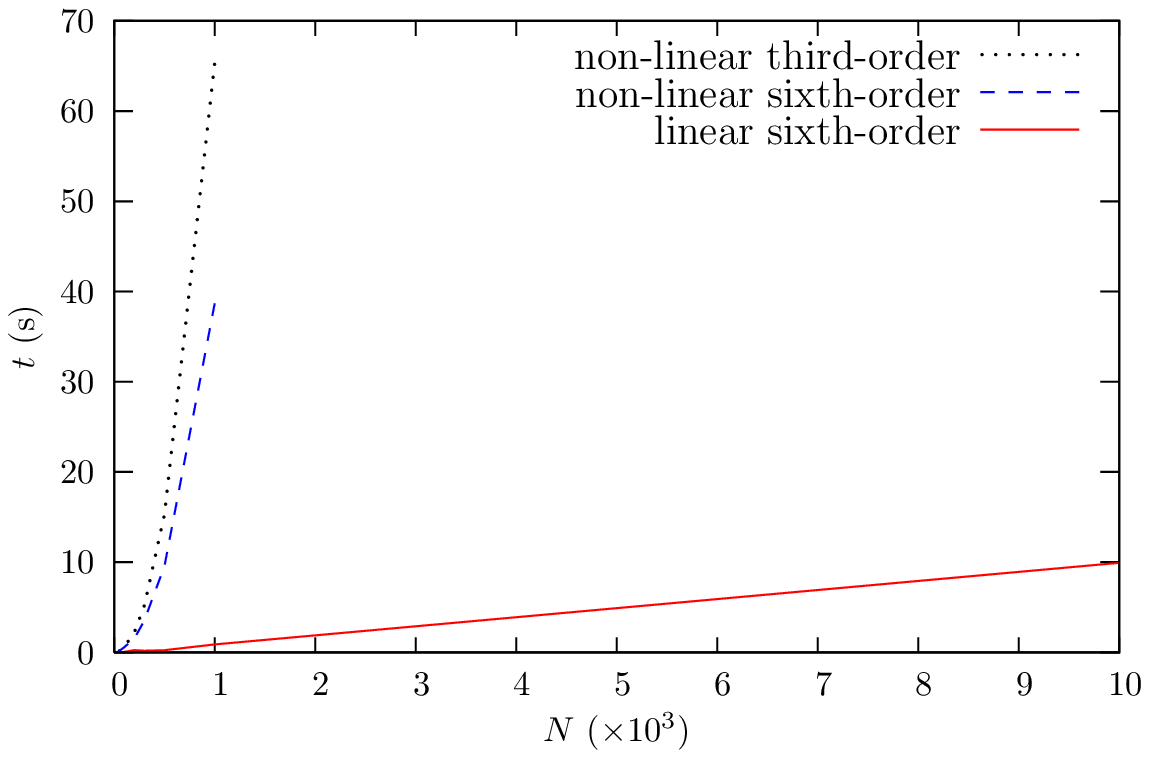}}
  \subfigure[Euclidean norms]{\label{f2}\includegraphics[scale=0.5]{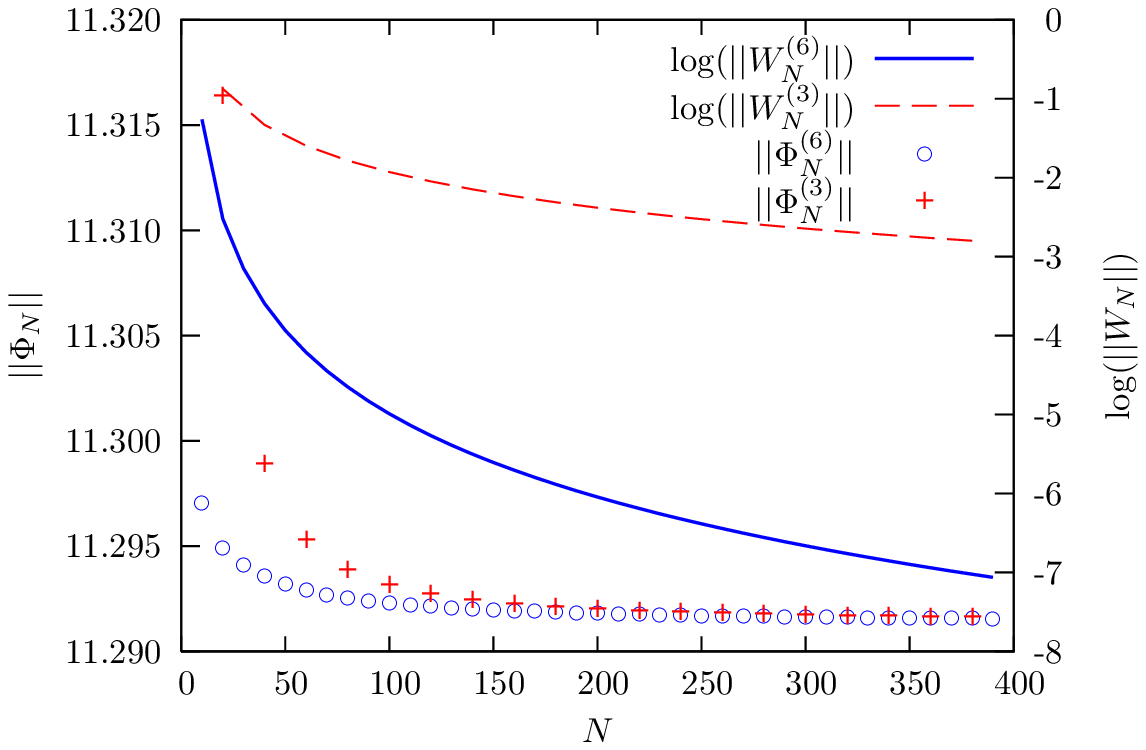}}
  \par\vspace{-0.4cm}
  \caption{Execution times (a) and exact ($||W_{N}||$) and mean
    ($||\Phi_{N}||$) Euclidean norms (b) as functions of the grid
    dimension $N+1$. The exact solution is given in
    \eqref{eq:sol_exata} and corresponds to the Gaussian density
    \eqref{eq:densidade} with boundary conditions $\phi_{1}=1$ and
    $\phi_{N+1}=2$}
  \label{fig:Tempos}  
\end{figure}

Figure~\ref{f1} shows the execution time as a function of the grid
dimension $N+1$ for three cases. In one case (the dotted line), the
numerical solution was computed by the non-linear third order
numerical scheme \eqref{eq:recurh}. In the second case (the dashed
line), the numerical solution was computed by the non-linear sixth
order numerical scheme \eqref{eq:recur}. In the last case (the solid
line), the numerical solution was computed by the linear sixth order
numerical scheme \eqref{eq:recur2}. At $N=1000$, the execution time of
the non-linear third (sixth) order numerical scheme is approximately
145 (51) times the execution time of the linear sixth order numerical
scheme. Clearly, we can see that the linearization process described
above plays an essential role in the present Numerov scheme.

In order to measure the accuracy of the present Numerov scheme, we
can compute the Euclidean norm
\begin{equation}
  \label{eq:Norma}
  ||W_{N}||=\sqrt{\sum_{i=1}^{N+1}\biggl[ \Phi^{(e)}(x_{i}) -
  \Phi^{(n)}_{i} \biggr]^{2}}
\end{equation}
where $\Phi^{(e)}$ stands for the exact solution \eqref{eq:sol_exata}
and $\Phi^{(n)}$ stands for the numerical solution. Figure~\ref{f2}
shows (right vertical axis) a comparasion between two Euclidean norms
\eqref{eq:Norma}: one (dashed line) using the third-order numerical
scheme \eqref{eq:recurh} and the other (solid line) using the
sixth-order numerical scheme \eqref{eq:recur2}. Note that, at $N=400$,
the exact Euclidean norm of the third-order scheme is approximately
four orders of magnitude above the exact Euclidean norm of the
sixth-order scheme.  Naturally, we can see that the sixth-order
numerical scheme \eqref{eq:recur2} is much more accurate and efficient
than the third-order numerical scheme \eqref{eq:recur}. Of course, we
don't know the exact solution in practical applications. In that case,
the best we can do is to compute the mean Euclidean norm of the
numerical solution $\Phi^{(n)}$,
\begin{equation}
  \label{eq:NormaMedia}
  ||\Phi_{N}||=\sqrt{\frac{1}{N}\sum_{i=1}^{N+1}\biggl( \Phi_{i}^{(n)}
    \biggr)^{2}}.
\end{equation}
This mean Euclidean norm can be used as a convergency criterion, as
shown in Figure~\ref{f2} (left vertical axis). 

\section{Conclusions}
\label{sec:conclus}
We have applied the Numerov method to derive a sixth-order numerical
scheme to solve the one-dimensional Poisson equation
\eqref{eq:Poisson1D} with Dirichlet boundary conditions. The resulting
recurrence relations were exactly solved and the corresponding
execution time was linearized [see \eqref{eq:recur2}] in such way to
avoid the handling of a dense matrix. Therefore, the numerical scheme
\eqref{eq:recur2} is both accurate and efficient as illustrated in
Figure~\ref{fig:Tempos}. Moreover, it is extremely ease to implement
in any numerical or algebraic computer language. As pointed by
J.~M.~Blatt~\cite{jcp1jmb1967}, the Numerov method is both a
three-point method, which implies it is stable, and of highest order,
which implies it is accurate. All these features make the numerical
scheme \eqref{eq:recur2} the canonical method of choice for the
integration of the Poisson equation \eqref{eq:Poisson1D}.

\section*{Acknowledgment}
The author wish to thank Rafael Casalverini for useful discussions and
FAPESP for financial supports.

\bibliographystyle{unsrt} 

\end{document}